\pdftrailerid{} 
\documentclass[conference]{IEEEtran} 
\IEEEoverridecommandlockouts
\usepackage{cite}
\usepackage{amsmath,amssymb,amsfonts}
\usepackage{algorithmic}
\usepackage{graphicx}
\usepackage{textcomp}
\usepackage{balance}
\usepackage{xcolor}
\usepackage{url}

\def\BibTeX{{\rm B\kern-.05em{\sc i\kern-.025em b}\kern-.08em
    T\kern-.1667em\lower.7ex\hbox{E}\kern-.125emX}}

\makeatletter
\def\footnoterule{\relax%
  \kern-5pt
  \hbox to \columnwidth{\hfill\vrule width 0.5\columnwidth height 0.4pt\hfill}
  \kern4.6pt}
\makeatother

\renewcommand{\IEEEkeywordsname}{Keywords}

\begin{document}

\title{Hypersparse Traffic Matrices from Suricata Network Flows using GraphBLAS
  \thanks{
Research was sponsored by the Department of the Air Force Artificial Intelligence Accelerator and was accomplished under Cooperative Agreement Number FA8750-19-2-1000. The views and conclusions contained in this document are those of the authors and should not be interpreted as representing the official policies, either expressed or implied, of the Department of the Air Force or the U.S. Government. The U.S. Government is authorized to reproduce and distribute reprints for Government purposes notwithstanding any copyright notation herein.
}
}

\author{\IEEEauthorblockN{
Michael Houle$^1$, Michael Jones$^1$, Dan Wallmeyer$^2$, Risa Brodeur$^2$, Justin Burr$^2$, \\ Hayden Jananthan$^1$, Sam Merrell$^2$, Peter Michaleas$^1$, Anthony Perez$^2$, Andrew Prout$^1$,  Jeremy Kepner$^1$
    \\
    \IEEEauthorblockA{
    $^1$MIT, $^2$Center for Internet Security, Inc.
    }}}

\maketitle

\IEEEtitleabstractindextext{
\begin{abstract}
  Hypersparse traffic matrices constructed from network packet source and destination addresses is a powerful tool for gaining insights into network traffic.  SuiteSparse:GraphBLAS, an open source package for building, manipulating, and analyzing large hypersparse matrices, is one approach to constructing these traffic matrices.  Suricata is a widely used  open source network intrusion detection software package.  This work demonstrates how Suricata network flow records can be used to efficiently construct hypersparse matrices using GraphBLAS.
\end{abstract}

}

\IEEEpeerreviewmaketitle
\IEEEdisplaynontitleabstractindextext

\section{Introduction}

Large scale analysis of hypersparse network traffic matrices has been shown to be useful in gaining insight into wide range of network activity~\cite{jones2022graphblas, trigg2022hypersparse}.   Traffic matrices provide significant compression of network data and it can be advantageous to constructed traffic matrices in network sensors. The scale of this collection is best addressed by promoting the ubiquitous deployment of as many network sensors as possible.  GraphBLAS~\cite{kepner16mathematical,GraphBLASOrg} is an open standard software library that is ideally suited for both constructing and analyzing anonymized hypersparse traffic matrices.  A significant advantage of traffic matrices is the ability to construct analysis that are independent of anonymization.  Thus, anonymization of network packet source and destination addresses (e.g., using CryptoPAN~\cite{fan2004prefix}) can be performed at the sensor and helps meet the requirements of data protection frameworks with minimal impact on subsequent traffic analysis.

\begin{figure}[ht]
  \center{\includegraphics[width=1.0\columnwidth]{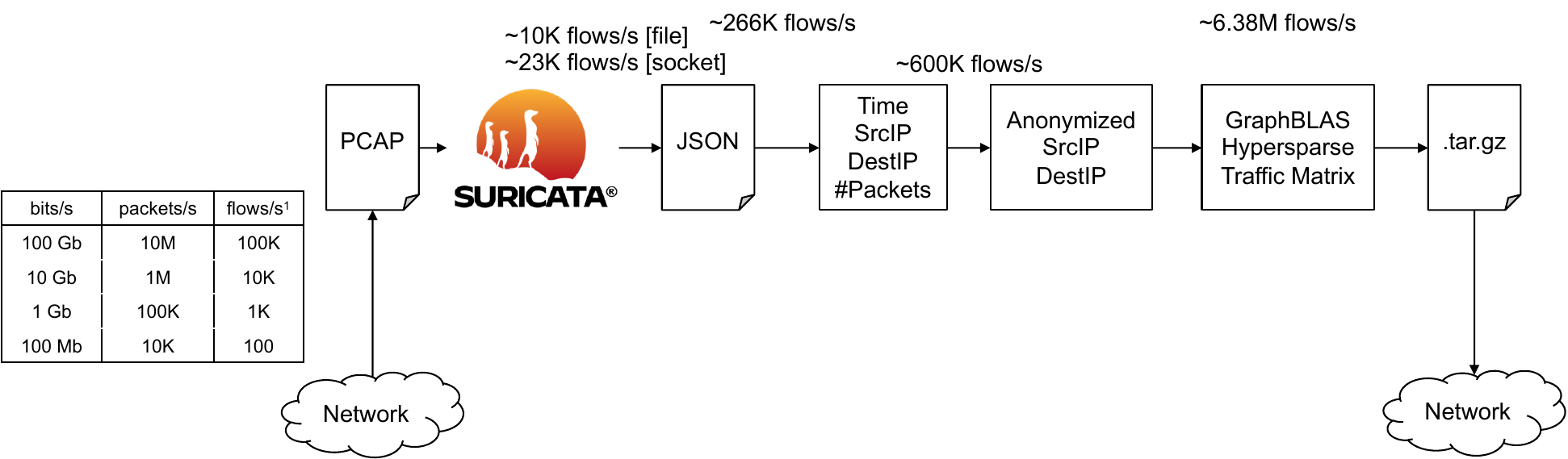}}
  \caption{{\bf Suricata Traffic Flow Processing}. A PCAP file containing CAIDA network telescope data is read in by the Suricata IDS in offline replay mode, and Suricata is configured to emit 'flow' records as JSON via an EVE logging target.  This JSON log file is then parsed by our program which converts Suricata flow records into anonymized GraphBLAS traffic matrices. $^1$Assumes 100 packets per flow for typical real-world traffic \cite{jurkiewicz2021flow}.}
  \label{fig:SuricataFlows}
\end{figure}

\section{Software Implementation}

Suricata~\cite{SuricataIDS} is a leading open source, high performance network traffic analysis and intrusion detection tool used by multiple vendors in their products.  Suricata performs traditional network packet inspection by organizing traffic into packet flows between two hosts, and can be configured to output this data as JSON formatted flow records to a log file or a unix domain socket.  Each flow record contains a source address, a destination address, and separate counts for packets flowing in either direction, plus additional fields not relevant for building traffic matrices.  Suricata can be configured to output flow records through multiple mechanisms, including through a software plug-in, a standard file system log file, or a unix domain socket.
The log file and unix domain socket options effectively serialize the flow records and allow for simpler ingest and processing.

SuiteSparse:GraphBLAS is a full implementation of the GraphBLAS standard, which defines a set of sparse matrix operations on an extended algebra of semirings using a wide range of operators and types.   A small utility program in C, {\tt json2grb}, was to developed to read and process Suricata flow records into GraphBLAS hypersparse matrices.  By default, the flow records are parsed to build matrices representing a window of $2^{17}$ packets, with groups of 64 of these matrices stored in a Unix TAR file for later collection and processing.  These sizes strike a good balance between granularity of matrices and files sizes for subsequent parallel processing.  Each GraphBLAS hypersparse matrix represents a ${2^{32}}\times{2^{32}}$ address space, using the IPv4 source and destination addresses of the network packets as indices.  Both the source and destination addresses can be anonymized using the CryptoPANT~\cite{cryptopANT}. The matrices are built from an array of index pairs indicating traffic flow direction, source to destination or destination to source, and the number of packets as indicated by the flow record.  The memory requirement for the source data used to build the matrix for these sizes is typically less than 1.5MB, and flushing each constructed matrix to disk ensures the memory requirement of the {\tt json2grb} utility remains low.  The GraphBLAS library contains a function to export hypersparse matrices in an LZ4 compressed format, minimizing the disk storage requirements.  Depending upon the nature of the traffic, a typical, compressed hypersparse matrix representing 2$^{17}$ packets is less than 420KB in size.  A Unix tar file of 64 hypersparse matrices is  less than 26MB.  Traffic dominated by a few flows produce much smaller files.

For our implementation tests, we used a small virtual machine with 4 virtual CPUs and 4 GB of memory that  is representative of lower-end of the resources allocated to a typical deployed network sensor.  The underlying physical CPUs were Intel(R) Xeon(R) Gold 6430 operating at 2.1GHz.  Suricata was installed locally on the network sensor, and a PCAP-format network packet capture file containing ~18 million packets was used to provide reliable and repeatable testing.  The packets were sourced from the largest public Internet observatory, the Center for Applied Internet Data Analysis (CAIDA) Telescope, that operates a variety of sensors including a continuous stream of unsolicited packets from a darkspace representing approximately 1/256 of the Internet.  The CAIDA traffic is almost entirely adversarial traffic and represents a near worst-case for traffic analysis with a large number of unique source and destination pairs.

\section{Test Environment and Experimental Results}

The flow record processing pipeline of our virtual test environment is illustrated in Figure 1.  Suricata can read source packets directly from the network or, for repeatable testing or deferred processing, from network packets captured in PCAP files.  For our tests, Suricata was configured to output the JSON flow records to a unix domain socket.  The {\tt json2grb} utility read records from the unix domain socket, and parsed them using the yyjson~\cite{yyjson} library.  The  addresses and packet flow counts were extracted from each flow record.  The addresses were then anonymized using the CryptoPAN library, and two records were added to the build array for the traffic matrix from each flow record, one for packets originating from the source address going to the destination address, and another for packets flowing from the destination address to the source address.  A count of the packets added to the build list was kept, and when it reached the predefined window size (default $2^{17}$), a GraphBLAS hypersparse matrix was created and written to the output TAR file.  A new TAR file was created after every 64 matrices written.

As can be seen from the sample values included in the figure, on our test virtual environment Suricata output $\sim$23,000 flow records per second.  This is likely related to how Suricata processes traffic flows and, more importantly how it determines when a particular flow ends, than the actual performance of our test environment.  The C JSON parsing library embedded in {\tt json2grb} could parse those flow records at about 10 times that rate ($\sim$266,000 flow records/second), while the anonymization occurred even quicker ($\sim$600,000 flow records/second).  The highly optimized SuiteSparse GraphBLAS routines performed even better, representing the smallest portion of the processing time, with $\sim$6.38 million flow records/second saved as traffic matrices.  The archived traffic matrix TAR files can then be retrieved from the sensor system at a later time and manner dependent on the actual sensor system and environment.

\section{Summary}

Ensuring broad deployment of traffic sensors is vital to enabling effective and informative analysis using GraphBLAS hypersparse matrices.  Adding traffic collection sensors to existing network tools helps to achieve this goal.  Suricata is a leading high-performance network traffic analysis and intrusion detection tool used by multiple vendors in their products.  This paper demonstrates how such a collection sensor can be added to a Suricata based system with minimal effort or operational impact.  Suricata natively supports outputting network packet flow records in an easily parsable JSON format that can be processed in a deferred manner through log files, or immediately via a unix domain socket.  Parsing, anonymizing and building the hypersparse traffic matrices can be completed faster than the native Suricata flow processing.  Data processing and storage requirements for the individual hypersparse matrices of $2^{17}$ packets is minimal.  The {\tt json2grb} utility written to demonstrate the capabilities for this paper never exceeded 512MB of system RAM usage while running.  Storage of the compressed hypersparse traffic matrices by GraphBLAS required less than 420KB per $2^{17}$ packets, or less than 26 MB for a full TAR file containing 64 matrices.

\section*{Acknowledgments}

The authors wish to acknowledge the following individuals: D. Anderson, L. Anderson, W. Arcand, S. Atkins, W. Bergeron, D. Bestor, C. Birardi, B. Bond, A. Bonn, S. Buckley, A. Buluc, D. Burrill, C. Byun, K Claffy, C. Conrad, T. Davis, C. Demchak, A. Edelman, G. Floyd, V. Gadepally, J. Gottschalk, T. Hardjono, C. Hill, M. Houle, M. Hubbell, C. Leiserson, P. Luszczek, K. Malvey, C. Milner, S. Mohindra, G. Morales, L. Milechin, J. Mullen, R. Patel, A. Pentland, H. Perry, S. Pisharody, C. Prothmann, A. Prout, S. Rejto, A. Reuther, J. Rountree, A. Rosa, D. Rus, M. Sherman, G. Wachman, S. Weed, C. Yee, M. Zissman.

\balance
\bibliographystyle{ieeetr}
\bibliography{hpec2024-suricata-graphblas}
\end{document}